# Gamma background and neutron flux measurements at Basic Ecological Observatory Moussala

*Y. Dimov, M. Guelev, L. Hristov, P.Ivanov, A. Mishev, A. Nishev, J. Stamenov and B. Vachev*

Institute for Nuclear Research and Nuclear Energy-Bulgarian Academy of Sciences
72 Tsarigradsko chaussee Sofia 1784 Bulgaria

*Abstract:*
*The Basic Ecological Observatory (BEO) Moussala is located at 2925m above sea level and one of the main activities is connected with gamma background and neutron dose rate monitoring.*
*The neutron and gamma measurements are carried out with help of site radiation monitor Harwell Type 3208-1. The dose rate of ambient gamma background is measured with SBN-90, improved with the embedded microprocessor system using automatic PC based system. A cross check and inter calibration with existing TLD (thermo luminescent dosimeters) at BEO Moussala which is a part of the monitoring network of INRNE is presented. A polyethylene sphere with a layer of lead serving as a neutron breeder, extending the energy range from about 40 MeV to several hundred MeV is presented as a project in development. Another project in development is the construction of active detector based on SNM-15 detectors for a dose rate measurement of cosmic neutrons is described. Preliminary study of the performances of the detectors are carried out. A further project for muonic telescope similar to the yet developed is discussed.*

## 1.Intoduction

The location of BEO Moussala is on the top of the highest mountain at Balkan peninsula. This is one of the most proper places in the region of Balkans and gives excellent possibility for high-mountain monitoring i.e. possibilities for measurements, experiments and monitoring for changes and processes at the atmosphere, pollution, air-transport, aerosol investigations, changes of gamma-background, dose-rate from neutron flux, cosmic ray investigations (Stamenov, Vachev 2003) etc… At the same time the analyses of collected data gives information about relation between very different kind of parameters and factors.

All the data are transmitted from BEO to INRNE trough Internet using the communication system developed system at the top (Ivanov et al. 2002a). The communication system gives the possibility to obtain information in quasi real time.

One of the most important measurements carried out at BEO Moussala are the radiation measurements precisely the gamma background and the obtained dose rate from neutrons. The observed radiation background at the top is superposition of radiation due to the specifics of the top such as soils, rocks etc…. The other contribution is due to the secondary cosmic ray radiation. In this paper we present the recent radiation measurement carried out at BEO Moussala and discuss the further improvements.

## 2. Quasi real-time Radioactivity Background Monitoring

The quasi real-time radioactivity background monitoring includes measurements for dose-rate of gamma and neutron flux. The gamma background monitoring is carried out by two different independent devices – SBN-90 and Harwell 3208-1. The Harwell 3208-1 provides also measurement of neutron flux. Both devices are connected with personal computers (PC), which give the possibility for full automatisation of the measurements. Additional measurement of the gamma background is carried out using thermo luminescent dosimeters (TLD) detector.

### 1.1 Gamma background real-time monitoring

The gamma background monitoring at BEO Moussala has been started in 2000. The French Electrical Company (EDF) consigned to INRNE four devices SBN-90 in 1994. One of the devices was based on top Moussala and the others are placed in different referent points – two are at INRNE, and one at Radioactive Waste Repository at Novi Khan. Several improvements are proposed and realized at INRNE. The original output of the device was changed with numerical output with help of the ADC developed at INRNE. At the same time software, which permits to manage the readout and data acquisition system is developed. This permits to collect the data at PC and to organize them in databank. The connection and the management of the device with PC and taking into account the existence of the communication system in BEO Moussala gives the possibility for a quasi real time to obtain information about the gamma background at the top. In 2002 CERN consigned to INRNE another one different type device – Harwell 3208-1 site radiation monitor which has been working on top Moussala since September 2002.

SBN-90 is based on scintillation detector of NaI(T). In this devices has been integrated microcontroller PIC16C77 and upgraded with new electronic block for amplification, reception and automatic recording the data from measurements (Stamenov et al. 2001). A method for calibration of the device was proposed and carried in the INRNE (Stamenov et al. 2001).

The average month dose-rate from gamma background on top Moussala for 2003 is shown on fig.1a. The minimal dose-rate is 87.50 nSv/h (May 2003) and the maximal is 128.10 nSv/h (July 2003). The floating of values and their fluctuations depends from rains. The measured dose-rate increases in months with intensive rains, because the rains collect the aerosols from higher level of the atmosphere and transport them to the ground-level. In the beginning of rain the dose-rate goes up and after that slowly decries. The recent results of the measurements are presented in fig. 1b, actually the obtained dose rate for mounth of August 2004. Comparing to the average dose-rate in Sofia as was expected the dose rate at the top is higher. In one this is due to the some specifics of BEO Moussala, the type of soils and rocks which are with Ra. On the other hand taking into account that the BEO Moussala is at about 3000m above the sea level the contribution of secondary cosmic ray radiation to the total gamma background is most important comparing to seal level.

The Site Radiation Monitor Harwell Type 3208-1 is low-level combined gamma and neutron radiation monitor. It has been designed to meet radiological protection requirements at the 400 GeV Proton Synchrotron at CERN, Lab. II Geneva. The monitor is constructed in environmental cases for outdoor usage and the design contains a total dose integration system capable of running for long periods except for routine battery replacements. The operation principle is as follows: after detection of each increment of integrated dose a pulse is produced by the electronic circuits. Afterwards the pulses are counted be a mechanical register. The gamma radiation measurement is carried out with help of argon filled ionisation chamber. The monitor provides pulse output after the accumulation of 50 nSv/h dosage. The device is connected to PC and registration of pulses into file is controlled by software developed by BEO-group (Ivanov et al. 2001b). The output pulse-signal is reduced to TTL.

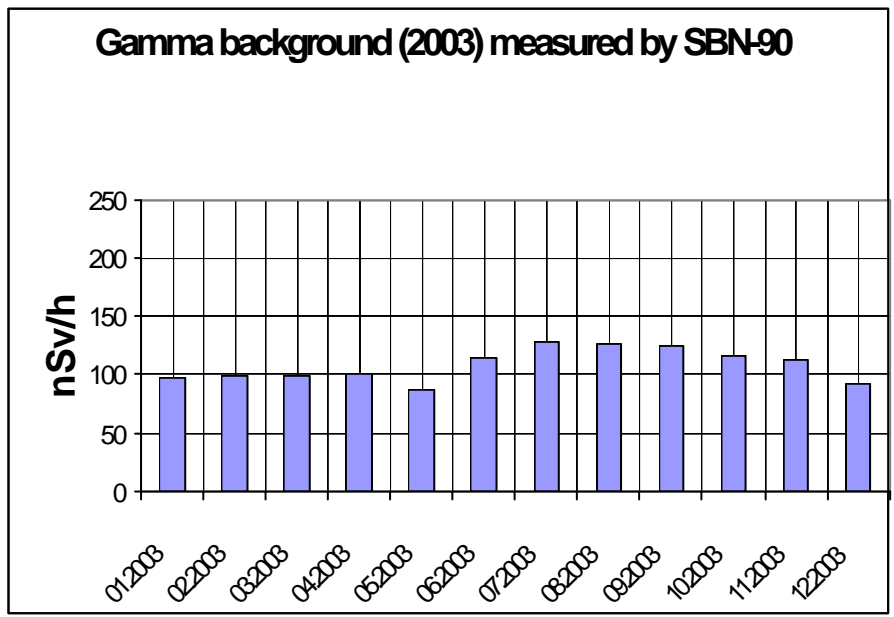

Fig.1a Average month dose-rate gamma background for 2003 on top Moussala measured by SBN-90

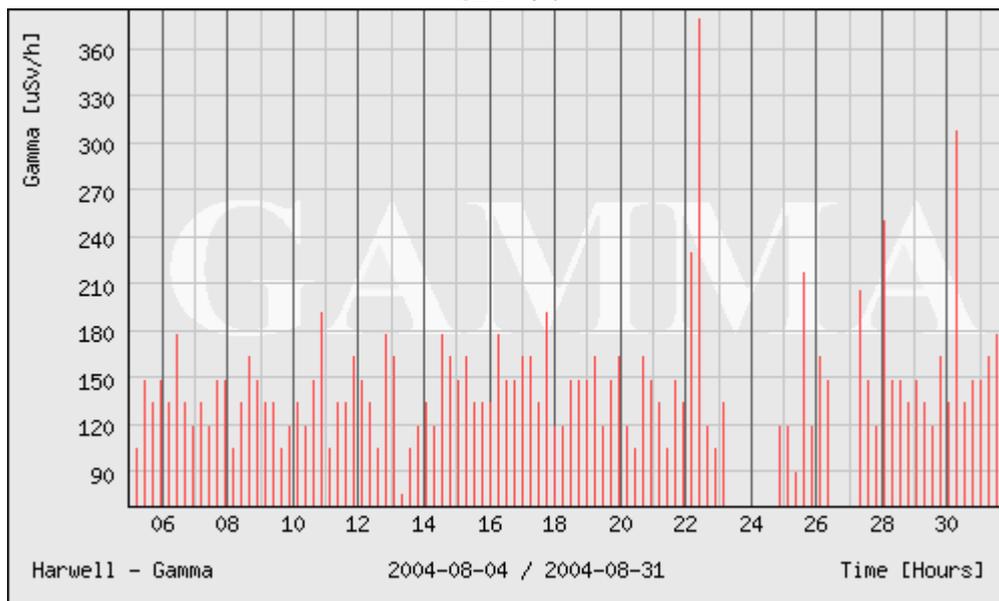

Fig.1b Average dose-rate gamma background for August 2004 on top Moussala measured by SBN-90 and Harwell 3208-1

The average dose rate measured by Harwell 3208-1 is 137.40 nSv/h (fig.2). Actually there are not systematical differences between values measured by SBN-90 and Harwell 3208-1. The differences are result of different sensitivity of devices and different kind of detectors.

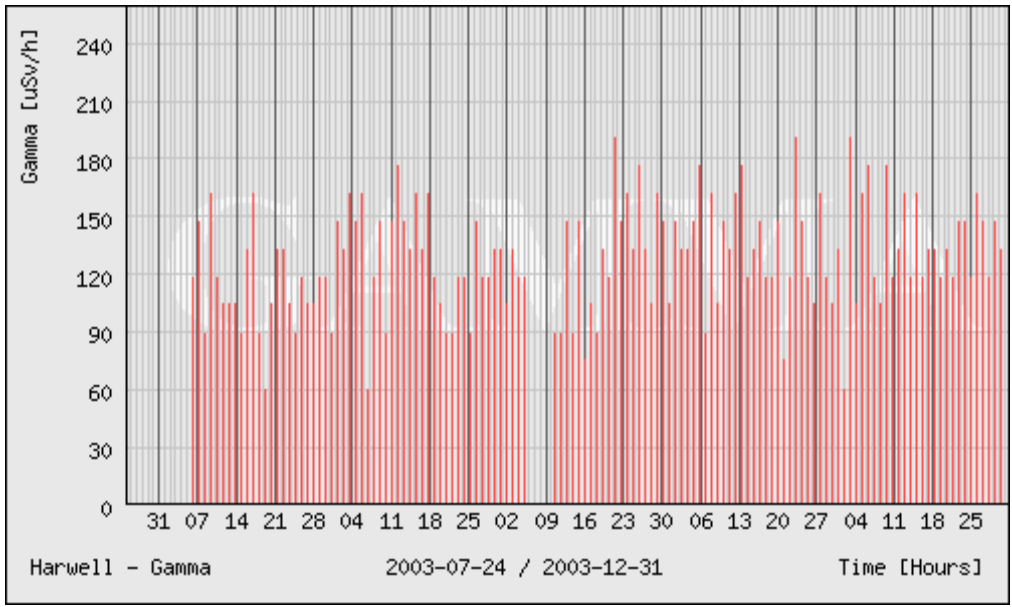

Fig.2 Average month dose-rate from gamma background (2003) measured by Harwell 3208-1

A TLD detector was also developed (Guelev et al. 1994). CaSO4:Dy – based TL dosimeter were developed and began use in Bulgaria during the 1980's. The dosimeter passed the requirements of IEC Standard 1066 in the early 1990's. Energy response of the dosimeter TLE-4, obtained using different linear combinations of detector response under tin and polyethylene filters: Sn – energy response of CaSO4:Dy detector placed in a 2 mm tin sphere; Pe - energy response of CaSO4:Dy detector placed in a 0.5 mm polyethylene holder. The energy and angular response responses of the TLD are presented at fig. 3a and fig. 3b. The experiment results are compared with Monte Carlo simulation (Isabey et al. 1997) carried out with EGS4 (Nelson et al. 1985) code. The annual average of the obtained dose-rate with the TLD for the same period as mentioned above is 160 nSv/h.

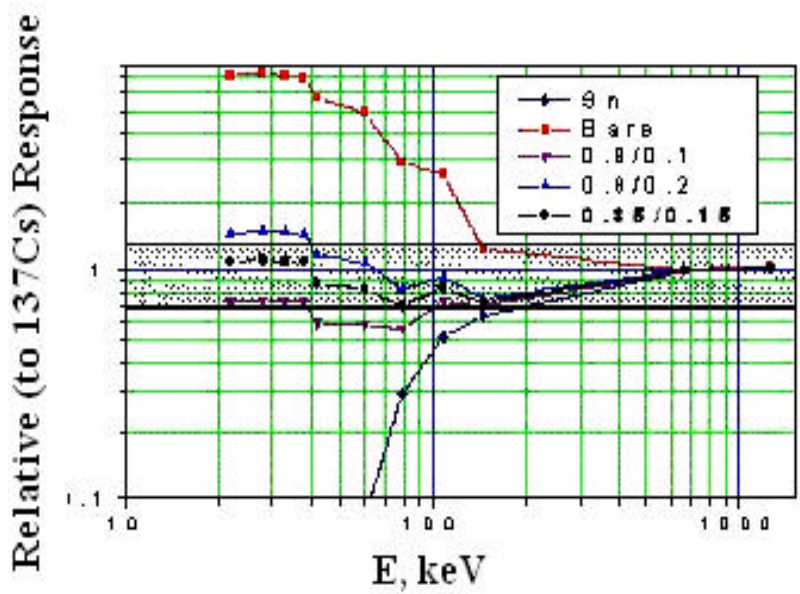

Fig. 3a The Energy response of the thermoluminescent dosimeter

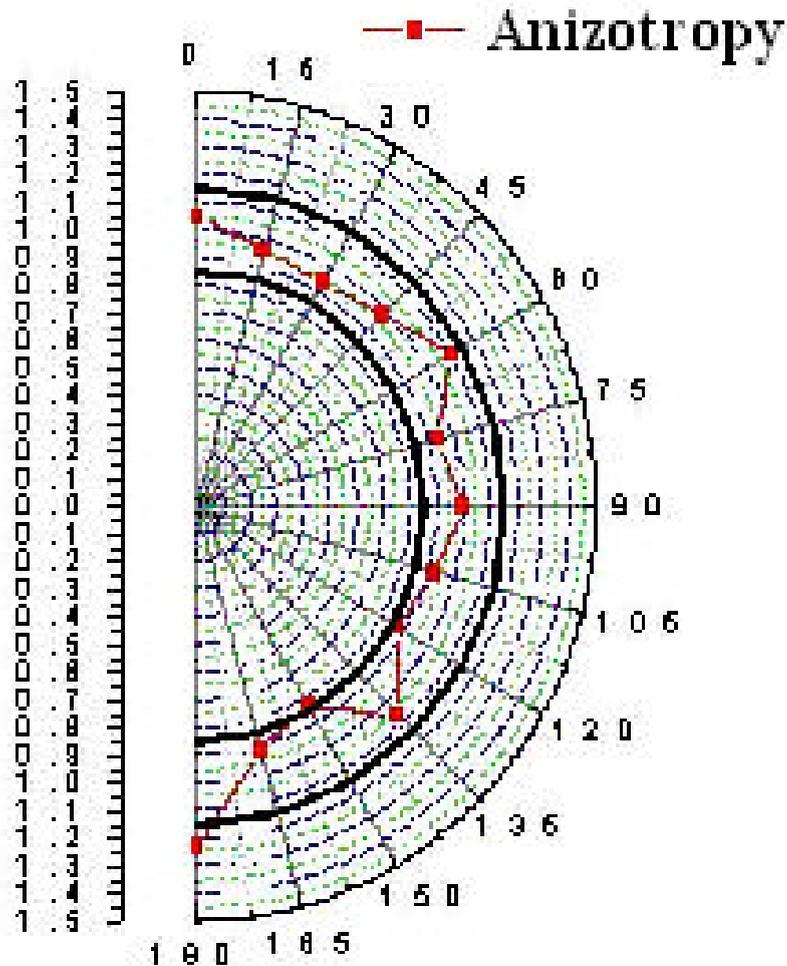

Fig. 3b The angular response of the thermoluminescent dosimeter

In all cases of measured gamma background there are not observed systematic errors. The differences are due essentially to the different sensitivity, energy range and methodology for dose-rate estimation.

**1.2 Neutron flux measurements**
The neutron flux on top Moussala has been measuring by Harwell 3208-1. The device use a proportional counter and Anderson-Braun moderator. The pulses from the counter are amplified and a trigger a pulse amplitude discriminator give a pulse output. The output pulses are counted in the neutron register and also added to the ionisation chamber pulses. The average month values for neutrons for 2003 is shown on fig.4a. The average annual dose-rate from neutron flux is 40.50 nSv/h. As example in fig. 4b are presented also the recent results for August 2004.

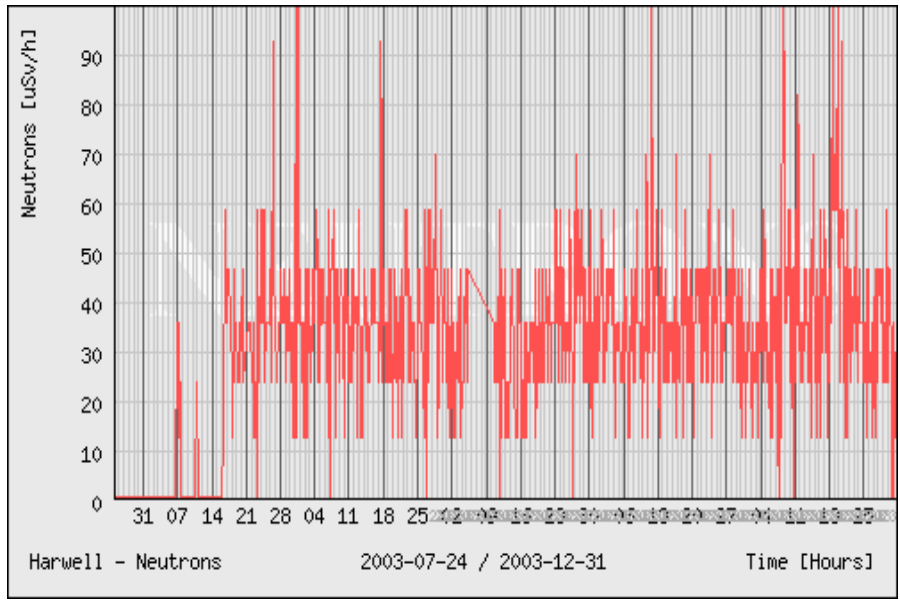

Fig.4a Average month values for neutrons

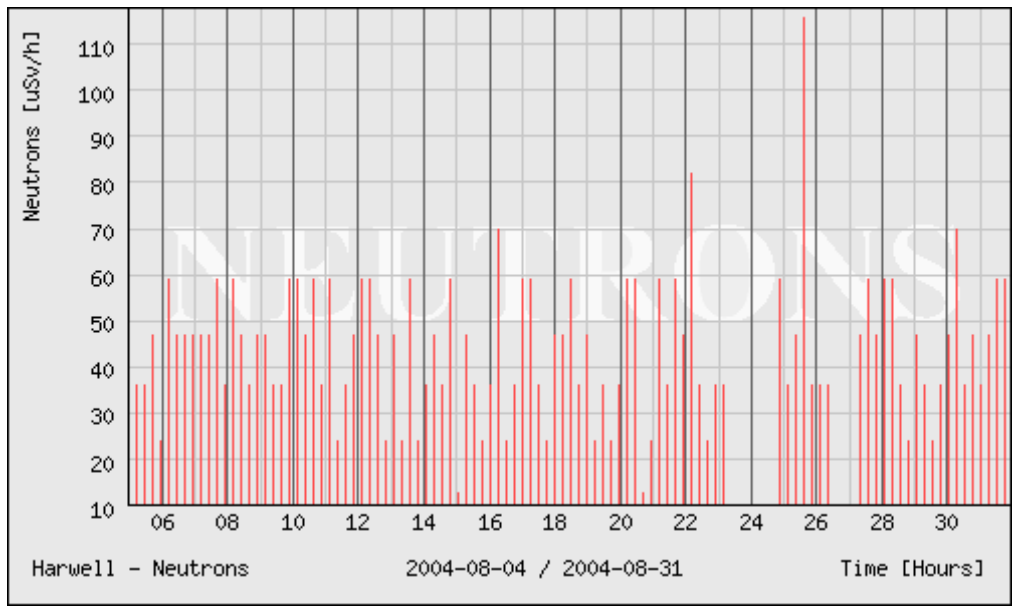

Fig.4b Average dose-rate from neutrons August 2004 measured by Harwell 3208-1

In a full analogy of the previous section the obtained dose-rate comparing to measurement at INRNE campus (500m above sea level) with the same device is bigger. This is due essentially to secondary cosmic ray radiation contribution.

**3 Future projects and plans**

One of the future projects yet in development is a neutron detector for absolute secondary cosmic neutron flux measurements. That kind of monitor will give good possibilities for analyzing the dose-rate of secondary neutron flux and average energy of integral spectra on secondary neutrons also. It will be based on Russian gas detectors filled with $BF_3$ of type SNM-15. The detectors will be situated under the roof of the building of BEO Moussala. On fig. 5 is shown the possible configuration and designs of the monitor for measurement of absolute neutron flux – group of 6 of detectors will be covered with polyethylene. The expected frequency of counting is 200 imp/sec for neutrons with energy from 1 MeV to 10 MeV.

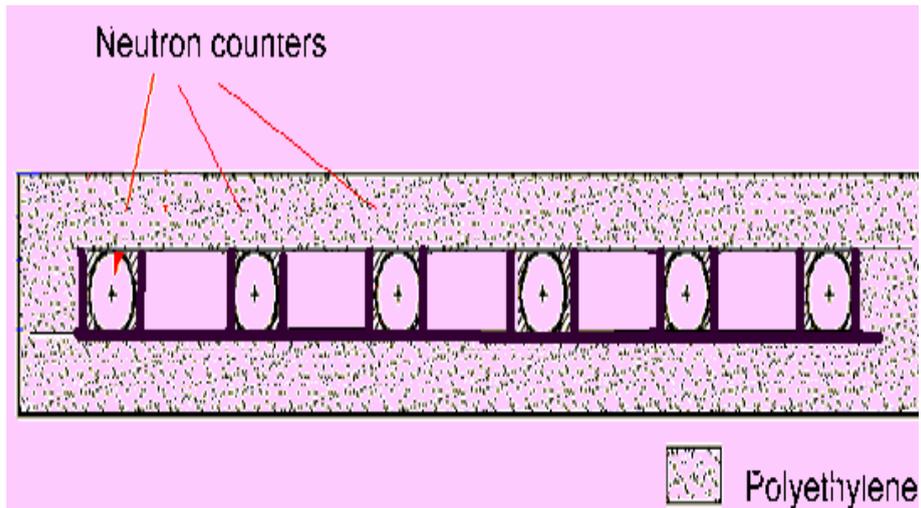

Fig. 5 Configuration of detectors of monitor for absolute neutron flux

This detector configuration is without lead i.e. without neutron multiplicator. This is the main difference comparing to the usual neutron monitors. Thus the principle aim of this device is the measurement of the absolute neutron flux of secondary cosmic ray radiation. It is clear that the precise Monte Carlo modeling of the detector response is needed, the aim being to estimate precisely the expected dose-rate.
A preliminary experimental study of the detector efficiency using neutron flux is carried out.

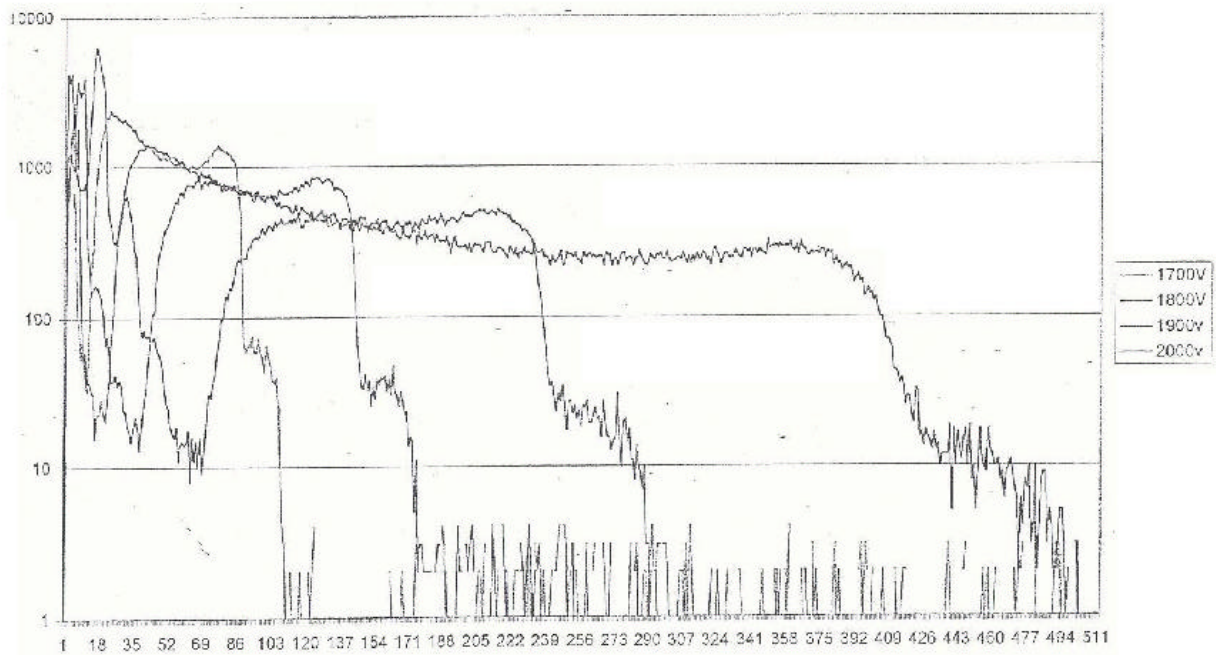

Fig. 6 Differential spectrum of SNM-15 obtained using neutron flux

Development of device for automatic real-time monitoring of absolute neutron flux will provide analyzes for secondary neutron flux, measurement of the dose-rate for aircraft crews and calibration of new detectors and devices (F. Spurny, I. Votoekova, 1993).

Under development in INRNE is a polyethylene sphere with a layer of lead serving as a neutron breeder, extending the energy range from about 40 MeV to several hundred MeV. We are also planning the construction of an active device, in which in the center of the sphere is placed a photodiode in contact with a $^{10}$B layer.

Importantt project in development is the commissioning of muon telescope based on water Cherenkov detectors. Such kind of device is similar to the yet builded one in South West University Neofit Rilski in Blagoevgrad Bulgaria (E. Malamova et al. 2001). The aim is measuring the muonic component of the secondary cosmic ray and afterwards searching correlation with other measurements as example weather changes, cloud cover and thunderstorms, in other words space weather.

## 4. Summary

We presented the gamma background measurements carried out at BEO Moussala. The gamma background dose-rate is obtained using different devices and methods. The differences of the obtained values are due essentially to the different sensitivity and range of the devices. The further development of active detector for neutron flux detection and afterwards the calculation of the dose rate obtained by cosmic ray neutrons will complete the radiation measurements at BEO Moussala as the development of the muonic telescope.

**Acknowledgements**

We warmly acknowledge I. Kalapov and L. Hristov for their work connected with SBN-90 and M. Vitkova for his assistance. We also acknowledge prof. F. Spurny for the fruitfull discussions and suggestions especially connected with neutron measurements. We are thankfull to N. Chikov for the assistance during the preliminary work with SNM-15 detectors. Finally we want to thank to Y. Dimov and E. Mihaylova for their work with TLD detectors and E. Dermendjiev and I. Ruskov for the suggestions concerning neutron detection.

**References:**

M. Guelev et al.,1994. Radiation Protection Dosimetry **51**, 25-4

R. Isabey et al., 1997.Nucl. Instr. Meth. **B 132**, 114-118

P. Ivanov et al., 2002a. Proc of **ELECTRONICS ET 2001** Sept. Sozopol Bulgaria v.3.,p58

P.Ivanov et al.,2002b. Proc of **ELECTRONICS ET 2002**, v4, 142-145, 26-28 Sept. Sozopol Bulgaria

E. Malamova et al., 2001. Proc. at **27 ICRC Hamburg 2001**, 3952-3955

W. Nelson, H. Hirayama, D.W.O. Rogers, 1985. The EGS4 code system, SLAC report 265, Stanford, CA94305

F. Spurny , I. Votoekova, 1993. Nucl. Ener. Safety **1** (39), 112-113, (1993)

J. Stamenov, B. Vachev, 2003. Proc. of **Himontonet European workshop** 28 June- 3 July 2003 Borovetz Bulgaria 48-80

J. Stamenov et al.,2001. Proc of **ELECTRONICS ET 2001** Sept. Sozopol Bulgaria, v. 2, p62